\documentclass{article}
\usepackage{spconf,amsmath,graphicx}
\usepackage{mathrsfs}
\usepackage{amsfonts}
\usepackage{bm}
\usepackage{url}
\usepackage{color}

\newcommand{\Ai}{\boldsymbol{A}_i}

\newcommand{\ain}{\boldsymbol{a}_{in}}

\newcommand{\bim}{\boldsymbol{b}_{im}}
\newcommand{\Bi}{\boldsymbol{B}_i}

\newcommand{\Dim}{\boldsymbol{D}_{im}}
\newcommand{\Em}{\boldsymbol{e}_m}

\newcommand{\eyeM}{\boldsymbol{E}_{M}}

\newcommand{\gimn}{\tilde{g}_{inm}}
\newcommand{\gimnd}{\tilde{g}_{in'm}}
\newcommand{\Gin}{\boldsymbol{G}_{in}}
\newcommand{\Gindiag}{\boldsymbol{\mathcal{G}}_{in}}

\newcommand{\Hp}{^\mathsf{H}}
\newcommand{\lambdaim}{\lambda_{im}}

\newcommand{\Qi}{\boldsymbol{Q}_{i}}

\newcommand{\qim}{\boldsymbol{q}_{im}}
\newcommand{\qimhat}{\hat{\boldsymbol{q}}_{im}}
\newcommand{\qimH}{\boldsymbol{q}^{\mathrm{H}}_{im}}
\newcommand{\qimhatH}{\hat{\boldsymbol{q}}^{\mathrm{H}}_{im}}
\newcommand{\rim}{r_{im}}
\newcommand{\rimhat}{\hat{r}_{im}}

\newcommand{\sigmaijn}{\sigma_{ij,n}}
\newcommand{\sij}{\bm{s}_{ij}}

\newcommand{\sumi}{\sum_i}
\newcommand{\sumijn}{\sum_{i,j,n}}
\newcommand{\sumj}{\sum_j}
\newcommand{\sumn}{\sum_n}

\newcommand{\sumk}{\sum_k}

\newcommand{\sumijm}{\sum_{i,j,m}}
\newcommand{\tik}{t_{ik}}
\newcommand{\tikd}{t_{ik'}}

\newcommand{\Tp}{^\mathsf{T}}
\newcommand{\uim}{\boldsymbol{u}_{im}}
\newcommand{\uimhat}{\hat{\boldsymbol{u}}_{im}}

\newcommand{\vkj}{v_{kj}}
\newcommand{\vkjd}{v_{k'j}}
\newcommand{\win}{\boldsymbol{w}_{in}}
\newcommand{\Wi}{\boldsymbol{W}_i}

\newcommand{\xij}{\boldsymbol{x}_{ij}}
\newcommand{\xijhat}{\boldsymbol{x}\Hp_{ij}}

\newcommand{\yij}{\bm{y}_{ij}}

\newcommand{\zkn}{z_{kn}}
\newcommand{\zknd}{z_{k'n'}}

\newcommand{\Hline}{\noalign{\hrule height 0.4mm}}

\title{Regularized fast multichannel nonnegative matrix factorization with ILRMA-based prior distribution of joint-diagonalization process}
\name{
\shortstack{Keigo Kamo$^1$ \qquad Yuki Kubo$^1$ \qquad Norihiro Takamune$^1$ \qquad Daichi Kitamura$^2$\\ Hiroshi Saruwatari$^1$ \qquad Yu Takahashi$^3$ \qquad Kazunobu Kondo$^3$}
\thanks{This work was partly supported by SECOM Science and Technology Foundation and JSPS KAKENHI Grant Numbers JP19H01116 and JP19K20306.}
}
\address{
$^1$The University of Tokyo, Tokyo, Japan \\
$^2$National Institute of Technology, Kagawa Collage, Kagawa, Japan \\ 
$^3$Yamaha Corporation, Shizuoka, Japan
}

\begin{document}
\ninept
\maketitle
\begin{abstract}
In this paper, we address a convolutive blind source separation (BSS) problem 
and propose a new extended framework of FastMNMF by introducing prior information 
for joint diagonalization of the spatial covariance matrix model. 
Recently, FastMNMF has been proposed as a fast version of multichannel nonnegative matrix factorization
under the assumption that the spatial covariance matrices of multiple sources can be jointly diagonalized.
However, its source-separation performance was not improved and the physical meaning of the joint-diagonalization process was unclear. 
To resolve these problems, we first reveal a close relationship 
between the joint-diagonalization process and the demixing system used 
in independent low-rank matrix analysis (ILRMA). 
Next, motivated by this fact, we propose a new regularized FastMNMF supported by ILRMA 
and derive convergence-guaranteed parameter update rules. From BSS experiments, 
we show that the proposed method outperforms the conventional FastMNMF in source-separation accuracy with almost the same computation time.
\end{abstract}
\section{Introduction}
\label{sec:intro}
Blind source separation (BSS)~\cite{sawada2019review} is a technique that separates sound sources from observed mixtures without any prior information about the sources or mixing system. 
For a determined or overdetermined situation, 
when the sources are point sources and reverberation is sufficiently short (referred to as  the \textit{rank-1 spatial model}),
frequency-domain independent component analysis~\cite{smaragdis1998blind,saruwatari2006blind}, independent vector analysis~\cite{hiroe2006solution,kim2006independent,kim2006blind}, and independent low-rank matrix analysis (ILRMA)~\cite{kitamura2016determined,kitamura2018determined}
have been proposed.
In particular, ILRMA is a BSS technique assuming statistical independence between the  sources and the low-rank structure of source spectrograms, 
and provides high-accuracy separation with a short computation time.
However, the rank-1 spatial model cannot hold in the case of spatially spread sources or strong reverberation.

Multichannel nonnegative matrix factorization (MNMF)~\cite{ozerov2010multichannel,sawada2013multichannel}
is an extension of nonnegative matrix factorization (NMF)~\cite{lee1999learning} to the multichannel case, which estimates the spatial covariance matrices of each source.
MNMF employs full-rank spatial covariance matrices~\cite{duong2010under} and this model can
simulate situations where, e.g., the reverberation is longer than the length of time-frequency analysis.
However, it has been reported that MNMF has a huge computational cost and its performance strongly depends on the initial values of parameters~\cite{kitamura2016determined}.
To accelerate the parameter estimation,
Ito and Nakatani have proposed \textit{FastMNMF}~\cite{ito2019fastmnmf}, which is an improved algorithm of MNMF under the assumption of jointly diagonalizable spatial covariance matrices. 
It has been reported that, although the computation time of the algorithm is greatly reduced, its source-separation performance is still sensitive to the parameter initialization 
and not always improved (indeed, it is almost the same as that of the original MNMF)~\cite{kubo2019efficient}.
In addition, the physical meaning of the joint-diagonalization process in FastMNMF is unclear; consequently, prior information cannot be introduced into the parameter optimization to achieve further improvement.

To resolve the above-mentioned problems, 
we provide three contributions in this paper, namely, 
a new FastMNMF framework with physically reasonable prior information,
its parameter optimization algorithm based on a new type of coordinate descent, 
and an experimental evaluation of the proposed FastMNMF. 
First, we reveal that the joint-diagonalization process in FastMNMF 
is closely related to the demixing system used in ILRMA. 
Motivated by this fact, we propose a new {\it regularized FastMNMF} with 
the prior distribution of the joint-diagonalization matrix supported by ILRMA. 
Next, we derive parameter update rules on the basis of vectorwise coordinate descent (VCD)~\cite{mitsui2018vectorwise} that guarantees a monotonic nonincrease in the cost function.
Finally, we conduct BSS experiments under reverberant conditions, 
showing that the proposed FastMNMF outperforms 
the conventional FastMNMF as well as ILRMA in source-separation accuracy while maintaining similar computational efficiency.
\section{Conventional Methods}
\label{sec:conventional}
\subsection{Formulation}
Let the numbers of sources and channels be $N$ and $M$, respectively.
The short-time Fourier transforms (STFTs) of the multichannel source, 
the observed signal, and the separated signal are defined as
\begin{align}
    \sij &= (s_{ij,1},\dots,s_{ij,N})\Tp \in \mathbb{C}^N, \\
    \xij &= (x_{ij,1},\dots,x_{ij,M})\Tp \in \mathbb{C}^M, \\
    \yij &= (y_{ij,1},\dots,y_{ij,N})\Tp \in \mathbb{C}^N,
\end{align}
where $i=1,\dots,I,\,j=1,\dots,J,n=1,\dots,N$, and $m=1,\dots,M,$ are the indices
of the frequency bins, time frames, sources, and channels, respectively, and $\cdot\Tp$ denotes the transpose.
\subsection{ILRMA~\cite{kitamura2016determined}}
When the window size in an STFT is sufficiently longer than the impulse responses between the sources and the microphones and the sources are point sources, we can represent the observed signal as
\begin{align}
\label{formula:mixing}
    \xij = \Ai\sij,
\end{align}
where $\Ai =(\boldsymbol{a}_{i1},\dots,\boldsymbol{a}_{iN})\in \mathbb{C}^{M\times N}$ is a 
frequency-wise mixing matrix and $\ain$ is the steering vector for the $n$th source.
If $M=N$ and the mixing matrix $\Ai$ is invertible, we can estimate the separated signal as
\begin{align}
\label{formula:demixing}
    \yij = \Wi\xij,
\end{align}
where $\Wi=(\bm{w}_{i1},\dots,\bm{w}_{iN})\Hp=\Ai^{-1}$ is the demixing matrix and $\cdot\Hp$ denotes the Hermitian transpose.
ILRMA assumes that the separated signals $y_{ij,n}$ ($n=1,\dots,N$) are statistically independent of each other, i.e.,
\begin{align}
\label{formula:independent}
    p(\yij) = \prod_n p(y_{ij,n}),
\end{align}
and each $y_{ij,n}$ follows the complex Gaussian distribution
whose mean is zero and variance is $r_{ij,n}$.
The source model $r_{ij,n}$ is a spectrogram of the $n$th source at the $i$th frequency and $j$th time frame, having a low-rank spectral structure represented by NMF.
From (\ref{formula:demixing}) and (\ref{formula:independent})，
the negative log-likelihood of the observed signal, which is a cost function to be minimized, is given by
\begin{align}
\label{cost:ILRMA}
    \mathcal{L}_{\mathrm{I}} \overset{c}{=} \sumijn\biggl[\frac{|\win\Hp\xij|^2}{r_{ij,n}}\!+\!\log r_{ij,n}\biggr] \!-\!2J\sumi\log|\det\Wi|,
\end{align}
where $\overset{c}{=}$ denotes equality up to a constant.
Since (\ref{cost:ILRMA}) w.r.t. the source model parameter $r_{ij,n}$ is the Itakura--Saito-divergence-based cost function, the parameter is updated by
the auxiliary function technique~\cite{hunter2000quantile}, similarly to Itakura--Saito NMF~\cite{fevotte2009nonnegative}.
Regarding the demixing matrix $\Wi$, the cost function (\ref{cost:ILRMA}) is the sum of the quadratic form of $\win$ and the negative log-determinant of $\Wi$.
This type of cost function can be minimized by iterative projection (IP)~\cite{ono2011stable}, which guarantees a monotonic nonincrease in the cost function.
The demixing matrix $\Wi$ can be optimized so as to make separated signals mutually independent. Details of these update rules are described in~\cite{kitamura2016determined}.
\subsection{FastMNMF~\cite{ito2019fastmnmf,sekiguchi2019fast}}
In convolutive BSS, the frequency-domain instantaneous mixing process is translated 
into a model using a rank-1 spatial covariance matrix $\ain\ain\Hp$ for each source.
In this case, the observed signal $\xij$ is modeled as follows:
\begin{align}
    \xij \sim \mathcal{N}(\boldsymbol{0}, \sumn r_{ij,n}\ain\ain\Hp).
\end{align}
A rank-1 spatial covariance model, however, is inappropriate
when reverberation is strong or the sources are not regarded as
point sources.
In the MNMF model, it is assumed that a spatial covariance matrix is full rank and denoted as $\Gin$ instead of the rank-1 spatial model $\ain\ain\Hp$.
Under this assumption, the observed signal is represented as
\begin{align}
\label{formula:MNMFobservation}
    \xij \sim \mathcal{N}(\boldsymbol{0}, \sumn\sigmaijn\Gin),
\end{align}
where $\sigmaijn$ is a source spectrogram.
It is also assumed that $\sigmaijn$ has a
low-rank structure, i.e.,
\begin{align}
\label{formula:sigma}
  \sigmaijn=\sumk\tik\vkj\zkn,
\end{align}
where $k=1,\dots,K$ is the index of the NMF basis, and 
$\tik\in\mathbb{R}_{\geq 0}$ and $\vkj\in\mathbb{R}_{\geq 0}$ represent the $i$th frequency component of the $k$th basis and the $j$th time-frame activation component of the $k$th basis, respectively.
In addition, $\zkn\in\mathbb{R}_{\geq 0}$ is a latent variable that indicates 
whether the $k$th basis belongs to the $n$th source.
In MNMF, we can estimate 
$\Gin$, $\tik$, $\vkj$, and $\zkn$
by minimizing the negative log-likelihood of $\xij$, but this
consumes a huge amount of computation.

To reduce the computational cost of the update algorithm, FastMNMF additionally assumes that the spatial covariance matrices
$\boldsymbol{G}_{i1},\dots,\boldsymbol{G}_{iN}$ are jointly diagonalizable
by $\Qi=(\boldsymbol{q}_{i1},\dots,$ $\boldsymbol{q}_{iM})\Hp$ as
\begin{align}
\label{formula:jointdiag}
    \begin{cases}
        \Qi\boldsymbol{G}_{i1}\Qi\Hp = \boldsymbol{\mathcal{G}}_{i1} \\
        \;\;\;\;\;\;\;\;\;\;\;\;\;\;\;\;\;\vdots \\
        \Qi\boldsymbol{G}_{iN}\Qi\Hp = \boldsymbol{\mathcal{G}}_{iN},
    \end{cases}
\end{align}
where $\Gindiag$ is a diagonal matrix.
From (\ref{formula:MNMFobservation}) and (\ref{formula:jointdiag}), 
the negative log-likelihood of the observed signal is given by
\begin{align}
\label{cost:FastMNMF}
    \mathcal{L}_{\mathrm{F}} &\overset{c}{=} \sumijm\biggl[\frac{|\qim\Hp\xij|^2}{\sum_{n,k}\tik\vkj\zkn\gimn} + \log\sum_{n,k}\tik\vkj\zkn\gimn\biggr] \nonumber\\
                &\quad -2J\sumi\log|\det\Qi|,
\end{align}
where $\gimn$ is the $m$th diagonal element of $\Gindiag$.
Similarly to ILRMA, $\Qi$ in (\ref{cost:FastMNMF}) can be optimized via IP and the remaining parameters
are updated by using the auxiliary function technique~\cite{sekiguchi2019fast}.
After the update, we can estimate the separated signals via the multichannel Wiener filter.
\section{Proposed Method}
\label{sec:proposed}
\subsection{Motivation and strategy}
\label{subsec:spatial}
The joint-diagonalization matrix $\Qi$ of FastMNMF makes the observed signal $\xij$ uncorrelated 
because $\xij$ follows the multivariate complex Gaussian distribution.
When we consider the rank-1 spatial model, the demixing matrix $\Wi$ in ILRMA is regarded as one of the decorrelation matrices.
From the definition of $\Wi$, the spatial covariance matrix $\ain\ain\Hp$ multiplied by the demixing matrix $\Wi$ on both sides becomes
\begin{align}
\label{formula:Wi_diagonize}
    \begin{cases}
        \Wi\boldsymbol{a}_{i1}\boldsymbol{a}_{i1}\Hp\Wi\Hp = \boldsymbol{e}_1 \boldsymbol{e}_1\Hp \\
        \hspace{22mm}\vdots \\
        \Wi\boldsymbol{a}_{iN}\boldsymbol{a}_{iN}\Hp\Wi\Hp = \boldsymbol{e}_N \boldsymbol{e}_N\Hp,
    \end{cases}
\end{align}
where $\boldsymbol{e}_n$ denotes the one-hot vector in which the $n$th element equals unity and the others are zero, and consequently $\boldsymbol{e}_n \boldsymbol{e}_n\Hp$ is a diagonal matrix.
Thus, this demixing matrix $\Wi$ is one of the joint-diagonalization matrices in the rank-1 spatial model.
On the other hand, when the spatial model is not rank-1, such as when the sources are still point sources but the 
reverberation is strong, 
the full-rank spatial covariance matrix $\tilde{\boldsymbol{G}}_{in}$ is defined 
as the sum of the covariances corresponding to the rank-1 part and the reverberation part $\sigma_{rev}\boldsymbol{\Psi}_{i}$~\cite{gustafsson2003source},
\begin{align}
\label{formula:SCM}
    \tilde{\boldsymbol{G}}_{in} =\ain\ain\Hp + \sigma_{rev}\boldsymbol{\Psi}_{i},
\end{align}
and the demixing matrix $\Wi$ can also jointly diagonalize the first term of the right-hand side of (\ref{formula:SCM}), as in (\ref{formula:Wi_diagonize}). 
Therefore, the joint-diagonalization matrix $\Qi$ can be approximated by
$\Wi$ estimated in ILRMA.
This fact motivates us to propose a new algorithm to find the optimal $\Qi$ around $\Wi$ that jointly diagonalizes rank-1 spatial covariance matrices.
Note that, although principal component analysis (PCA) is also a typical method of decorrelation,
the rotation matrix of PCA only diagonalizes the spatial covariance matrix of the observed signal $\xij$,
which is the weighted sum of the spatial covariance matrix $\Gin$ of each source,
but does not jointly diagonalize each one.
Thus, PCA is not appropriate for the joint diagonalization.

In this paper, we only consider a determined situation ($M=N$).
If $M<N$, i.e., underdetermined situations, the demixing matrix $\Wi$ cannot strictly diagonalize 
the first term of the right-hand side of (\ref{formula:SCM}).
However, we can still apply this method in this case because the demixing matrix $\Wi$ leads to the separated signals being independent of each other to some extent, i.e., $\Wi\Gin\Wi\Hp\,(n=1,\dots,N)$ is close to a diagonal matrix. 
\subsection{Proposed regularized FastMNMF}
\label{subsec:regularized}
From the discussion in Sec.~\ref{subsec:spatial},
we can introduce the prior distribution of the joint-diagonalization matrix $\Qi$ into (\ref{cost:FastMNMF}), where the mean of the distribution is set to the demixing matrix $\Wi$ of ILRMA, as
\begin{align}
\label{formula:qim}
    \qim &\sim \mathcal{N}(\qimhat, (J\lambdaim)^{-1}\eyeM),\\
\label{formula:qim=wim}
    \qimhat &= \boldsymbol{w}_{im}, 
\end{align}
where 
$\lambda_{im}$ is the weight parameter, $\eyeM$ is the $M\times M$ identity matrix, and
$J$ is used to remove the dependence on the total number of time frames.
Introduction of the prior distribution (\ref{formula:qim}) is equivalent to the imposition of the regularization term $J\sum_{i,m}\lambda_{im}||\qim-\qimhat||^2$ on (\ref{cost:FastMNMF}).
Hence, the negative log-posterior of the proposed regularized FastMNMF is obtained as
\begin{align}
\label{formula:reFastMNMF1}
    \mathcal{L}_{\mathrm{R}} &\overset{c}{=} \sumijm\biggl[\frac{|\qimH\xij|^2}{\sum_{n,k}\tik\vkj\zkn\gimn}+ \log\sum_{n,k}\tik\vkj\zkn\gimn\biggr] \nonumber \\
                &\quad -2J\sumi\log|\det\Qi| + J\sum_{i,m}\lambda_{im}||\qim-\qimhat||^2.
\end{align}

First, we derive update rules of the joint-diagonalization matrix $\Qi$.
We gather only the terms depending on $\qim$ in (\ref{formula:reFastMNMF1}) 
and rewrite the cost function as
\begin{align}
\label{formula:reFastMNMF2}
    \mathcal{L}_{\mathrm{R}} 
                &\overset{c}{=} J\sum_{i,m}\qim\Hp\Dim\qim -2J\sumi\log|\det\Qi| \nonumber\\ 
                &\quad - J\sum_{i,m}\lambda_{im}(\qimH\qimhat+\qimhatH\qim),
\end{align}
where 
\begin{align}
    \Dim = \frac{1}{J}\sumj\frac{\xij\xijhat}{\sum_{n,k}\tik\vkj\zkn\gimn}+\lambda_{im}\eyeM.
\end{align}
Equation (\ref{formula:reFastMNMF2}) is the sum of the quadratic form of $\qim$, the negative log-determinant of $\Qi$, and the \textit{linear terms} of $\qim$.
This type of problem cannot be solved by IP because of the existence of the linear terms. Instead of IP, VCD, which
we previously proposed~\cite{mitsui2018vectorwise}, can minimize (\ref{formula:reFastMNMF2}) w.r.t. $\qim$,
guaranteeing a monotonic nonincrease in the cost function.
We expand the term $\det\Qi$ in (\ref{formula:reFastMNMF2}) using $\Bi=(\boldsymbol{b}_{i1},\dots,\boldsymbol{b}_{iM})$,
which is the adjugate matrix of $\Qi$, defined as
\begin{align}
    [\Bi]_{mm'}=(-1)^{m+m'} \breve{\boldsymbol{Q}}_{i,m'm},
\end{align}
where $[\Bi]_{mm'}$ is the $(m,m')$th element of $\Bi$ and $\breve{\boldsymbol{Q}}_{i,m'm}$ is the $(m',m)$th minor determinant of $\Qi$. From a property of cofactor expansion, we obtain $|\det\Qi|^2=|\qim\Hp\bim|^2=\qim\Hp\bim\bim\Hp\qim$.
Note that $\bim$ is independent of $\qim$ from its definition~\cite{strang1993introduction}.
Therefore, the derivative of (\ref{formula:reFastMNMF2}) is obtained as
\begin{align}
    \frac{1}{J}\frac{\partial\mathcal{L}_{\mathrm{R}}}{\partial\qim^*}=\Dim\qim-\frac{\bim}{\qim\Hp\bim}-\lambdaim\qimhat,
\end{align}
where $\cdot^*$ denotes the complex conjugate.
By solving the equation $\partial\mathcal{L}_{\mathrm{R}}/\partial\qim^*=0$,
we describe the update rules of $\qim$ based on VCD as follows:
\begin{align}
\label{update:VCD1}
 \uim      &\leftarrow (\Qi\Dim)^{-1}\Em, \\
 \uimhat   &\leftarrow \lambdaim\Dim^{-1}\qimhat, \\
 \rim      &\leftarrow \uim^\mathrm{H}\Dim\uim, \\
 \rimhat   &\leftarrow \uim^\mathrm{H}\Dim\uimhat, \\
\label{update:VCD5}
 \qim     &\leftarrow
                \left\{
                    \begin{array}{ll}
                        \hspace{-2mm}\frac{\uim}{\sqrt{\rim}} +\uimhat,\hspace{75pt}(\mathrm{if}\,\rimhat=0)\\
                        \hspace{-2mm}\frac{\rimhat}{2\rim}\!\biggl[\!\sqrt{1\!+\!\frac{4\rim}{|\rimhat|^2}}\!-\!1\biggr]\uim \!+\!\uimhat \hspace{1.3mm}(\mathrm{otherwise}).
                    \end{array}
                \right.
\end{align}

Next, we describe update rules of the other parameters $\tik$, $\vkj$, $\zkn$, and  $\gimn$.
The cost function $\mathcal{L}_{\mathrm{R}}$ in (\ref{formula:reFastMNMF1}) w.r.t. $\tik$, $\vkj$, $\zkn$, and $\gimn$ is the same as $\mathcal{L}_{\mathrm{F}}$ in (\ref{cost:FastMNMF}) 
because the regularization term is a function of $\qim$ and independent of these parameters.
Then, the update rules are given in~\cite{sekiguchi2019fast} as
\begin{align}
 \tik &\leftarrow \tik\sqrt{\frac{\sum_{j,n,m}\frac{|\qimH\xij|^2\vkj\zkn\gimn}{(\sum_{k',n'}\tikd\vkjd\zknd\gimnd)^2}}{\sum_{j,n,m}\frac{\vkj\zkn\gimn}{\sum_{k',n'}\tikd\vkjd\zknd\gimnd}}}, \\
 \vkj &\leftarrow \vkj\sqrt{\frac{\sum_{i,n,m}\frac{|\qimH\xij|^2\tik\zkn\gimn}{(\sum_{k',n'}\tikd\vkjd\zknd\gimnd)^2}}{\sum_{i,n,m}\frac{\tik\zkn\gimn}{\sum_{k',n'}\tikd\vkjd\zknd\gimnd}}}, \\
 \zkn &\leftarrow \zkn\sqrt{\frac{\sum_{i,j,m}\frac{|\qimH\xij|^2\tik\vkj\gimn}{(\sum_{k',n'}\tikd\vkjd\zknd\gimnd)^2}}{\sum_{i,j,m} \frac{\tik\vkj\gimn}{\sum_{k',n'}\tikd\vkjd\zknd\gimnd}}}, \\
\label{update:ginm}
 \gimn &\leftarrow \gimn\sqrt{\frac{\sum_{j,k} \frac{|\qimH\xij|^2\tik\vkj\zkn}{(\sum_{k',n'}\tikd\vkjd\zknd\gimnd)^2} }{\sum_{j,k}\frac{\tik\vkj\zkn}{\sum_{k',n'}\tikd\vkjd\zknd\gimnd}}}.
\end{align}
These also use the auxiliary function technique, which guarantees a monotonic nonincrease in the cost function.
When $\lambdaim=0$, the update rules (\ref{update:VCD1}) to (\ref{update:ginm}) are the same as those of the conventional FastMNMF.
\subsection{Scheduling of weight parameter of regularizer}
\label{sec:annealing}
The demixing matrix $\Wi$ is not an accurate solution of FastMNMF because we cannot ignore the second term of the right-hand side of (\ref{formula:SCM}) in the full-rank spatial covariance case. Therefore, 
the weight parameter of the regularizer, $\lambdaim$, should become smaller in the latter part of the iterations and this annealing-like approach improves the separation accuracy.
\section{EXPERIMENT}
\subsection{Experimental conditions}
\label{sec:condition}
\begin{figure}[tb]
\begin{center}
 \includegraphics[width=8.5cm]{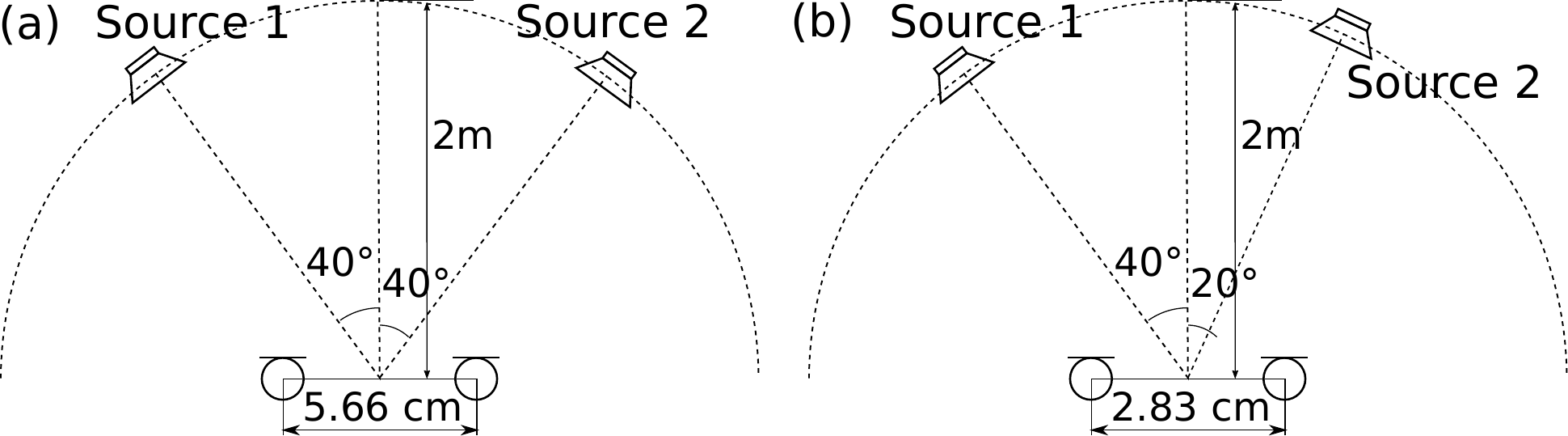}
\end{center}
\vspace{-5mm}
\caption{Spatial arrangements of sources and microphones.\label{mic}}
\vspace{-5mm}
\end{figure}

We confirmed the efficacy of the proposed method by conducting 
music source separation experiments.
We compared six methods: 
ILRMA~\cite{kitamura2016determined},
the conventional FastMNMF with $\eyeM$ initialization for $\Qi$ ({\bf FastMNMF w/ $\eyeM$ init.})~\cite{ito2019fastmnmf},
the conventional FastMNMF with PCA initialization for $\Qi$ ({\bf FastMNMF w/ PCA init.})~\cite{sekiguchi2019fast},
FastMNMF with $\Wi$ initialization for $\Qi$ ({\bf FastMNMF w/ $\Wi$ init.}) as a reference,
the proposed regularized FastMNMF without weight scheduling ({\bf proposed regularized FastMNMF 1}), and 
the proposed regularized FastMNMF with weight scheduling ({\bf proposed regularized FastMNMF 2}).
We used monaural dry music sources
of four melody parts~\cite{kitamura2018open}.
Eight combinations of instruments with different melody parts were selected as shown in Table~\ref{drySources}.
\begin{table}[t]
\vspace{-2mm}
\caption{Dry sources used in experiment\label{drySources}}
\begin{center}
\begin{tabular}{c|c|c}
\Hline
        & Part name & Source (1/2) \\ \hline
Music 1 & Midrange/Melody 2 & Piano/Flute \\
Music 2 & Melody 1/Melody 2 & Oboe/Flute \\
Music 3 & Melody 2/Midrange & Violin/Harpsichord \\
Music 4 & Melody 2/Bass     & Violin/Cello \\
Music 5 & Melody 1/Bass     & Oboe/Cello \\
Music 6 & Melody 2/Melody 1 & Violin/Trumpet \\
Music 7 & Bass/Melody 2     & Bassoon/Flute \\
Music 8 & Bass/Melody 1     & Bassoon/Trumpet \\
\Hline
\end{tabular}%
\end{center}
\vspace{-2mm}
\end{table}
To simulate reverberant mixing, the two-channel mixed signals were produced by convoluting the impulse response E2A ($T_{60}=300\,\mathrm{ms}$) in the RWCP database~\cite{nakamura2000acoustical}.
Fig.~\ref{mic} shows the recording conditions of E2A used in our experiments.
In these mixtures, the input signal-to-noise ratio was 0~dB.
The sampling frequency was 16~kHz and an STFT was performed using a 64 ms Hamming window with a 16 ms shift ($T_{60}$ is longer than the window length, i.e., the spatial covariance
matrices are full rank).
The total number of bases in the low-rank source model was $K=20$.
The initializations of the source model parameters ($\tik,\vkj,\zkn$) and the spatial covariance matrix $\Gin$ in FastMNMF 
were random values and the identity matrix, respectively. 
The initialization of $\Qi$ in the proposed methods was the identity matrix.
The weight parameter of the proposed regularized FastMNMF~1 was set to $10^{-7}$ and
that of the proposed regularized FastMNMF~2 in the $l$th iteration was set to 
$\lambdaim(l)=\lambda_0(\lambda_{end}/\lambda_0)^{l/L}$, where $L$ is the total number of iterations, $\lambda_0$ is $10^{-6}$, and $\lambda_{end}$ is $10^{-13}$. The number of iterations in the proposed and conventional methods was 300 and
that of ILRMA conducted before the proposed methods was 50.
We used the source-to-distortion ratio (SDR) improvement~\cite{vincent2006performance} to evaluate the 
total separation performance.

\begin{figure}[tb]
\begin{center}
 \includegraphics[width=8.5cm]{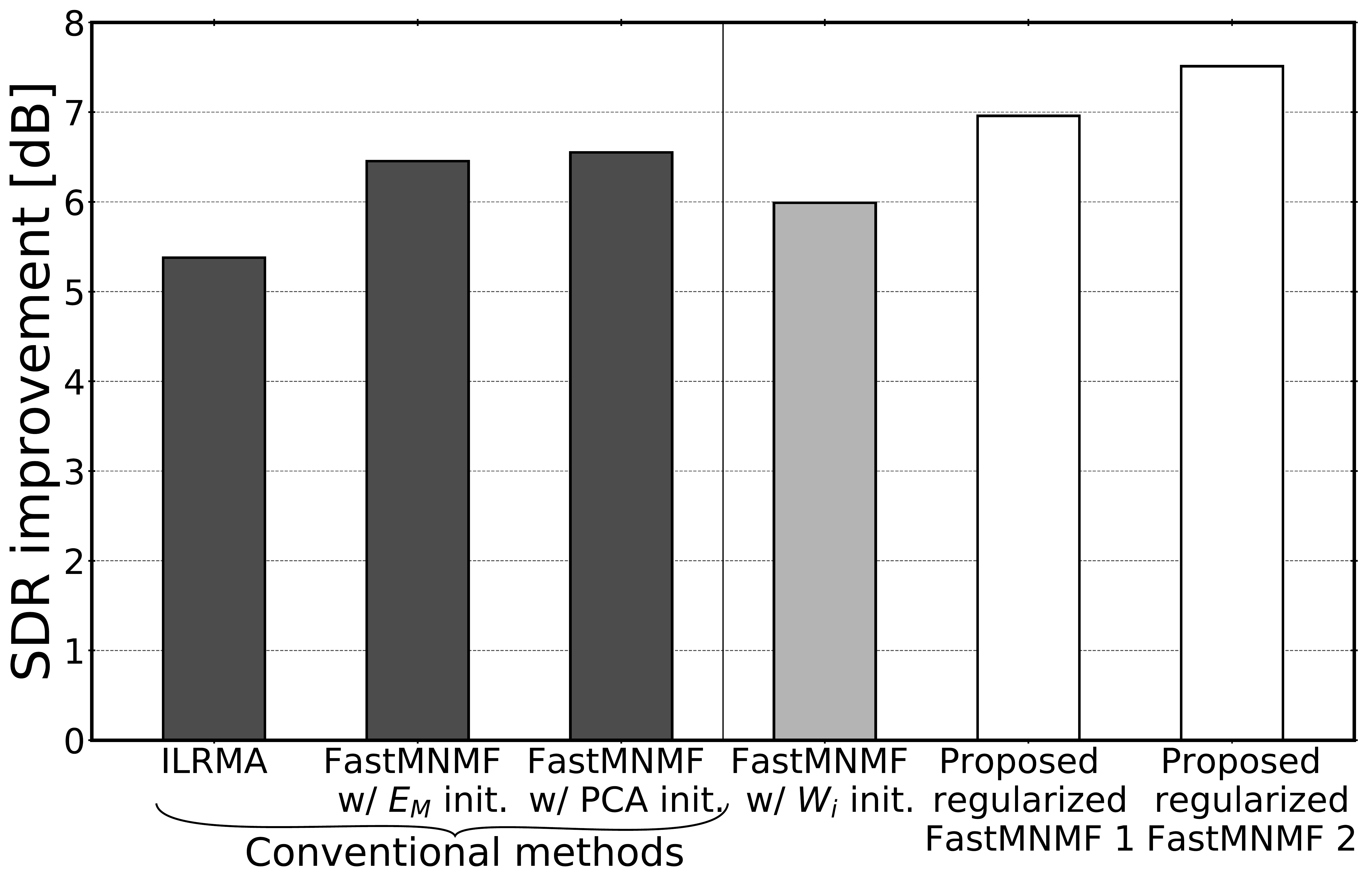}
\end{center}
\vspace{-5mm}
\caption{Resultant SDR improvement for each method. \label{result}}
\end{figure}
\vspace{-1mm}
\subsection{Experimental results for source-separation accuracy}
\vspace{-1mm}
Fig.~\ref{result} shows the average SDR improvements over the recording conditions, the source pairs, and 10-trial initialization. 
Compared with ILRMA, conventional FastMNMF w/ $\eyeM$ init. and FastMNMF w/ PCA init. provide 
better SDR improvements to some extent. The SDR improvement of FastMNMF w/ $\Wi$ init. is slightly lower than those of the conventional methods. 
On the other hand, the proposed regularized FastMNMF~1 and regularized FastMNMF~2 markedly outperform the conventional FastMNMF methods and ILRMA. This suggests that the initialization of $\Qi$ with $\Wi$ is not sufficient, showing the importance of introducing the new prior distribution for $\Qi$.

In addition, the proposed regularized FastMNMF~2 outperforms the proposed regularized FastMNMF~1.
This is because the joint-diagonalization matrix $\Qi$ of the proposed regularized FastMNMF~1 is exceedingly restricted by the demixing matrix $\Wi$ in ILRMA in the latter part of the iterations, which does not provide the best separation result as described in Sec.~\ref{sec:annealing}.

\vspace{-1mm}
\subsection{Experimental results for computation time}
\begin{figure}[tb]
\begin{center}
 \includegraphics[width=8.5cm]{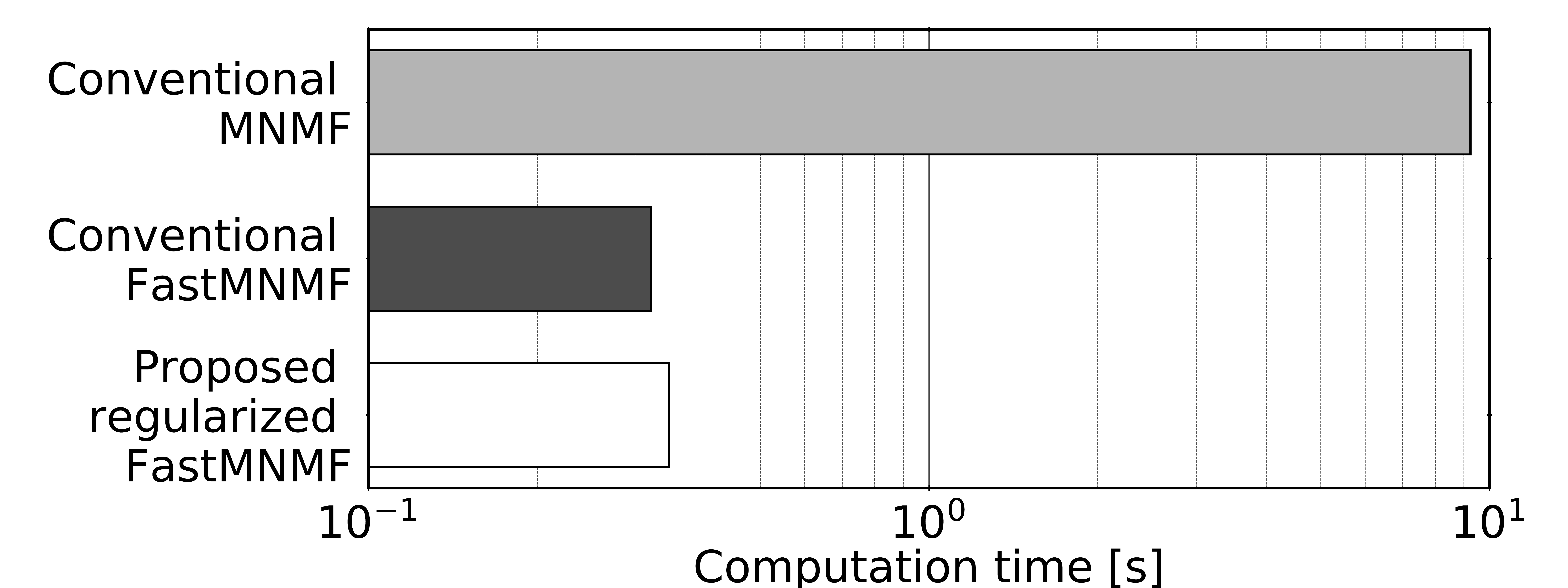}
\end{center}
\vspace{-5mm}
\caption{Average computation time per iteration for ``Music 1''.\label{time_compair}}
\end{figure}

\vspace{-1mm}
We measured the average computation time per iteration for ``Music~1''. We compared three methods: the conventional MNMF~\cite{sawada2013multichannel}, 
the conventional FastMNMF w/ $\eyeM$ init., and the proposed regularized FastMNMF~2.
Fig.~\ref{time_compair} shows that the 
proposed regularized FastMNMF and the conventional FastMNMF are much faster than the conventional MNMF.
The proposed FastMNMF~2 is slightly slower than the conventional FastMNMF
because of the VCD update (\ref{update:VCD1}) to (\ref{update:VCD5}),
but the difference is not significant.

%
%
%
%
%
\section{Conclusion}
In this paper, we first revealed that the joint-diagonalization matrix $\Qi$ in FastMNMF is closely related to the demixing matrix $\Wi$ in ILRMA.
Next, motivated by this fact, 
we proposed a new regularized FastMNMF that includes the prior distribution of $\Qi$ augmented with $\Wi$. 
Also, we derived the parameter update rules of $\Qi$ on the basis of VCD that guarantees a monotonic nonincrease in the cost function.
From the source-separation experiments, we showed that
the proposed method outperformed the conventional FastMNMF methods in SDR improvement with almost the same computation time.
\label{sec:conclusion}

\vfill\pagebreak

\bibliographystyle{IEEEbib}
\bibliography{strings,refs}

\end{document}